\documentclass[11pt,a4paper]{article}

\usepackage[margin=2.8cm,bottom=3.5cm]{geometry}

\usepackage{amsmath, amssymb}
\usepackage{color}
\pdfoutput=1 

\usepackage{graphicx} 
\usepackage{subcaption}
\usepackage[T1]{fontenc}

\title{Quasinormal modes in kink excitations and kink-antikink interactions: a toy model}

\author{
        Jo\~ao G. F. Campos \\
        Departamento de F\'isica, Universidade Federal de Pernambuco,\\
        Av. Prof. Moraes Rego, 1235, Recife - PE - 50670-901, Brazil\\
        jgfc@df.ufpe.br
            \and
        Azadeh Mohammadi\\
        Departamento de F\'isica, Universidade Federal de Pernambuco,\\
        Av. Prof. Moraes Rego, 1235, Recife - PE - 50670-901, Brazil\\
        azadeh.mohammadi@df.ufpe.br
}

\begin{document} 
\maketitle

\begin{abstract}
We study excitations and collisions of kinks in a scalar field theory where the potential has two minima with $Z_2$ symmetry. The field potential is designed to create a square well potential in the stability equation of the kink excitations. The stability equation is analogous to the Schr\"{o}dinger equation, and therefore we use quantum mechanics techniques to study the system. We modify the square well potential continuously, which allows the excitation to tunnel and consequently turns the normal modes of the kink into quasinormal modes. We study the effect of this transition, leading to energy leak, on isolated kink excitations. Finally, we investigate kink-antikink collisions and the resulting scaling and fractal structure of the resonance windows considering both normal and quasinormal modes and compare the results.
\end{abstract}

\section{Introduction}

Solitons and solitary waves are important solutions of nonlinear field equations that appear in many areas of physics including particle physics, condensed matter physics, cosmology and optics \cite{rajaraman1982solitons,vachaspati2006kinks,abdullaev2014optical}. They are particle-like solutions that have localized energy and propagate without losing the shape. Examples of solitons are vortices in superconductors \cite{abrikosov2004nobel,auslaender2009mechanics,fetter2018theory} and fluids \cite{kleckner2013creation} as well as cosmic strings \cite{vilenkin2000cosmic,hindmarsh2016new}, monopoles \cite{t1974magnetic,csaki2018kaluza} and instantons \cite{schaefer2002instanton,polyakov2018nucleon,schneider2018discrete} in high energy physics. There are also skyrmions \cite{fert2017magnetic} and domain walls \cite{parkin2008magnetic,koyama2011observation} in magnetic materials with interesting technological applications. Strictly speaking, solitons are solutions that collide elastically retaining only a phase after the collision. Example of such a system is the sine-Gordon model where the complete integrability of the model guarantees that it behaves like a true soliton. On the other hand, for most solitary waves there is no complete integrability of the model and thus we observe richer phenomena instead of just a phase shift after the collision.

In field equations in (1+1) dimensions there are solitary wave solutions called kinks which interpolate between neighboring minima of the potential. One can explain the behavior of many physical systems using the concept of solitary waves in one spatial dimension for instance in polyacetylene \cite{su1979solitons}, josephson junctions \cite{ustinov1998solitons} and propagation of light in optical fiber \cite{radhakrishnan1995bright} as well as in particle physics interactions \cite{dutta2008creating}. 
Also kinks can describe some higher dimensional structures, such as domain walls, which can be effectively viewed as one-dimensional structures in space. An important model that is well studied is the $\phi^4$ theory where the potential has $Z_2$ symmetry and two minima. In this model the collision between a kink and an antikink exhibit a rich behavior: they annihilate for small velocities and reflect for velocities larger than a critical velocity. Moreover, there are velocity intervals called resonance windows where the kinks bounce multiple times before escaping. These resonance windows, which occur due to exchange of the kink translational energy with the energy of bound normal mode excitations, alternate between intervals where the kink and antikink annihilate \cite{campbell1983resonance}.

The role of the normal mode excitations in kink-antikink collision has been already known since the work of Sugiyama \cite{sugiyama1979kink}, where the author could estimate the critical velocity of the $\phi^4$ using a collective coordinate approximation. The role of the normal modes excitation in the resonance windows was first explained in a seminal paper by Campbell et al. \cite{campbell1983resonance} where they showed that there is an interplay between the translational mode of the kink and the vibrational or "shape" mode in a resonant energy exchange mechanism. These computations were repeated by the same group for modified sine-Gordon models showing the effectiveness of their analysis \cite{peyrard1983kink, campbell1986kink}. In \cite{campbell1986kink}, Campbell et al. were able to observe self-similar structure in the resonance windows pattern of the so called double sine-Gordon model. Later, unequivocal evidence of fractal structure in the $\phi^4$ theory was given in \cite{anninos1991fractal}, where it was shown that the resonance windows form a nested structre where at the edge of a two-bounce windows there is a sequence of three-bounce windows, while at the edge of three-bounce resonance windows there is a sequence of four-bounce windows and so on. After that many authors investigated and contributed to understand the resonance phenomenon, as for instance in \cite{gani1999kink,goodman2005kink,goodman2007chaotic}.

Resonance windows are found in many other kink and antikink scattering models including recently the $\phi^4$ model variants \cite{simas2016suppression,bazeia2018scattering,gomes2018false}, the $\phi^6$ model and variants \cite{gani2014kink,bazeia2019kink,demirkaya2017kink}, the $\phi^8$ model \cite{gani2015kink,belendryasova2019scattering}, coupled two component kinks \cite{alonso2018reflection, halavanau2012resonance} and other models \cite{gani2018scattering}. In all these works the presence of vibrational modes is an essential ingredient to find resonance windows. For instance, in a model that interpolates between a degenerate vacuum with a vibrational mode to a vacuumless model without a vibrational mode, the resonance windows are suppressed as the vibrational modes disappear \cite{simas2017degenerate}. Surprisingly, resonance windows are also found in the $\phi^6$ model where the kinks do not exhibit normal modes. However, in this case the resonance is due to the interplay of the translational energy with the normal mode of the kink antikink pair \cite{dorey2011kink}. Similar results were found recently in a model where the kinks have power-law asymptotics \cite{belendryasova2019scattering}. Particularly, the study of collisions of kinks in such systems with long range tails is very interesting and requires specialized methods, as shown in \cite{manton2018forces,christov2019kink,christov2019long}. In some of these studies, approximate methods such as collective coordinate approximation have been used to partially understand the mechanism behind the resonance.  However, it is known that in many models the detailed pattern of the resonance windows cannot be captured by the collective coordinate approach. A recent approximate description of the resonance in the $\phi^4$ and $\phi^6$ models using the collective coordinates has been given in \cite{takyi2016collective,weigel2019collective}, where the authors discuss the applicability and the discrepancies between this approach and the exact results from the field equations. The studies in \cite{takyi2016collective,weigel2019collective} give evidence to the fact that the energy exchange mechanism of Campbell et al. is only an approximate description and it alone is not sufficient to create the resonance windows. In particular, it was pointed out in \cite{adam2019spectral} that this failure might happen because the normal modes disappear into the continuum during a kink-antikink collision.

Resonance windows are not limited to a kink colliding with an antikink. Interestingly, resonance windows are also observed in interactions of the kink with impurities and defects, where the energy exchange mechanism occurs between the translational mode of the kink and the impurity mode, as well as the kink vibrational mode, when it exists \cite{kivshar1991resonant, fei1992resonant,  fei1992resonant2, malomed1992perturbative, goodman2004interaction}. There can also be collision of kinks from different sectors \cite{gani2014kink}, multikink collisions \cite{marjaneh2017multi,marjaneh2017high,marjaneh2018extreme} or collisions of kinks with boundaries \cite{dorey2017boundary}, all having very interesting physics. Moreover, the inverse proccess where a kink and an antikink are created from particles and radiation has also been studied \cite{manton1997kinks,romanczukiewicz2006creation,dutta2008creating,romanczukiewicz2010oscillon}. Regarding applications, the $\phi^4$ model has been used recently to describe graphene deformations \cite{yamaletdinov2017kinks}. Furthermore, effects such as negative radiation pressure in kinks \cite{forgacs2008negative} were used to explain the absence of topological defects in the Universe today \cite{romanczukiewicz2017could}. Also in cosmology, higher dimensional solitons have been used to describe the universe \cite{rubakov1983we} and the collision of these solitons to describe the big bang \cite{khoury2001ekpyrotic}. 

In a model of highly interactive kinks where the tails of the kink decay as a power-law it was found that the kink has no vibrational mode, but has a quasinormal state \cite{gomes2012highly}. In a recent work, it was shown for a modified $\phi^4$ model that when the normal mode of the kink becomes a quasinormal mode (QNM) the resonance windows are suppressed \cite{dorey2018resonant}. In the present paper, we are also interested in investigating how isolated kink excitations and also collisions of a kink and an antikink are affected as the normal mode turns gradually into a QNM, using a toy model. For this to happen we look at the small fluctuations around the kink solutions and the corresponding Schr\"{o}dinger-like stability equations. 
We approximate the linear stability potential with a potential well allowing for normal modes and also a potential well with two barriers where the normal mode of the former turns into a QNM with energy leak. This gives us the opportunity to use quantum mechanical analytical techniques to find the normal modes of the system. Besides that, with this approximation we can study qualitatively a large class of models with similar linear stability potentials in the context of kink excitations and kink-antikink collisions considering QNMs. To have the stability potentials with desired shapes, we design two models with potential terms quadratic in the scalar field $\phi$, piecewise defined, with two minima. The potentials are constructed the way that they have kink solutions and the small fluctuations around these kinks have the aforementioned shapes in the stability potential, leading to normal modes and QNMs as we planned. In a completely different context, similar approach has been adopted in \cite{bordag2003spontaneous}, where the authors start with reflectionless linearized stability potentials and find their corresponding kink solutions and potentials.

Interestingly, although being artificially designed, the kink itself and kink-antikink collisions in our model have the expected features of the known solitons like kink of $\phi^4$ theory. We show that, in the model with the barriers, due to the energy leak, the fractal resonance windows are suppressed. Besides that, the critical velocity increases as a normal mode turns into a quasinormal one, matching the result in \cite{dorey2018resonant}.

The structure of the paper is as follows. In section \ref{model} we present the models we designed to have kink solutions. In section \ref{stability} we study the stability equation of small perturbations around the kink configurations as well as the effects associated to the transition between the normal and QNMs. In section \ref{collision} we compute the collision between a kink and an antikink and how it is affected when the QNMs appear. Finally, in section \ref{conclusion} we summarize our main results and conclusions.

\section{Model}
\label{model}

\begin{figure}[tbp]
\centering
  \includegraphics[width=0.76\columnwidth]{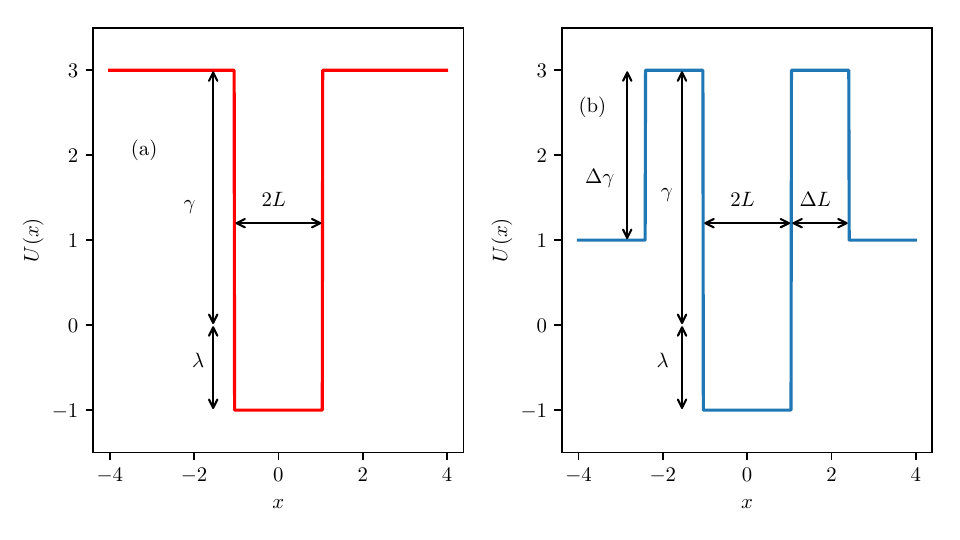}
        \caption{Linearized potentials $U(x)$ and $\tilde U(x)$. Parameters are: (a) $\gamma=3.0$, $\lambda=1.0$ and $L=1.047$. (b) $\gamma=3.0$, $\Delta\gamma=2.0$, $\lambda=1.0$, $L=1.045$, $\Delta L=1.365$.}
       \label{linearized_potential}
\end{figure}

Consider a scalar field theory in (1+1) dimensions given by the following Lagrangian
\begin{equation}
\mathcal{L}=\frac{1}{2}(\partial_\mu\phi)(\partial^\mu\phi)-V(\phi).
\label{lagrangian}
\end{equation}
Suppose that the system has $Z_2$ symmetry meaning $V(-\phi)=V(\phi)$ with two minima at $\phi=\pm\phi_0$ resulting in nontrivial kink and antikink solutions $\pm\phi_K(x)$. 
The stability of the kink can be studied by computing the behavior of small oscillations around a kink configuration. As one knows, if we write $\phi(x,t)=\phi_K(x)+\eta(x)e^{i\omega t}$, where $\eta$ is small, we find a Schr\"{o}dinger-like equation for $\eta$ as
\begin{equation}
H\eta\equiv \left[-\frac{d^2}{dx^2}+\frac{d^2V(\phi)}{d\phi^2}\bigg\rvert_{\phi=\phi_K}\right]\eta=\omega^2\eta.
\label{schrodinger}
\end{equation}
Let us define the linearized stability potential $U(x)\equiv \frac{d^2V(\phi)}{d\phi^2}\bigg\rvert_{\phi=\phi_K}$, in the following form
\begin{equation}
U(x)= 
  \begin{cases} 
     -\lambda, & 0\leq x<L, \\
     \gamma, & x>L,
      \end{cases}
      \label{stab-potential1}
\end{equation}
for positive $x$ and with the symmetry $U(x)=U(-x)$. In the above definition $\lambda$, $\gamma$ and $L$ are constant parameters.
This leads to a potential well, shown in Fig.~\ref{linearized_potential}(a), in the Schr\"{o}dinger-like equation (\ref{schrodinger}) with energies $\omega^2$ similar to the one appearing in quantum mechanics. As a result, the spectrum is known analytically. Now, let us modify the linear stability potential as
\begin{equation}
\tilde U(x)= 
  \begin{cases} 
     -\lambda, & 0<x<L, \\
     \gamma, & L<x<L+\Delta L, \\
     \gamma-\Delta\gamma, & x>L+\Delta L,
  \end{cases}
  \label{stab-potential2}
\end{equation}
with new constants $\Delta L$ and $\Delta\gamma$. We define $\gamma^\prime\equiv\gamma-\Delta\gamma$. This new potential gives the same potential well as before except for the additional barriers shown in Fig.~\ref{linearized_potential}(b). In this case one can also relate it to the corresponding problem in quantum mechanics and find the spectrum analytically. In the limiting case $\gamma' \to \gamma$ ($\Delta\gamma \to 0$) it is possible to recover the solutions in the Schr\"{o}dinger-like equation with a potential well originally designed in eq.\ (\ref{stab-potential1}). At this point, we plan to design the potential $V(\phi)$ in the Lagrangian (\ref{lagrangian}) with soliton solutions where the linear stability potentials of the kink excitations are the ones defined in equations (\ref{stab-potential1}) and (\ref{stab-potential2}). In \cite{dorey2018resonant}, the authors studied a kink of a deformed $\phi^4$ model with the stability potential which is close in shape to our potential well in Fig.~\ref{linearized_potential}(b).
Their model converges to the known $\phi^4$ model when a controlling parameter, $\epsilon$, is taken to zero. In this limit, the linear stability potential could be simplified to our $U(x)$ in Fig.~\ref{linearized_potential}(a).
The authors studied the kink-antikink interactions when the normal modes of the latter model ($\epsilon=0$) turn into the QNMs of the former one ($\epsilon\neq 0$). This "volcano shaped" linear stability potentials also appear in other models such as vacuumless systems \cite{bazeia2017sine} and kinks with power-law asymptotics \cite{gomes2012highly}. In the present work with our generic simplified stability potentials we plan to capture the essential aspects of the kink excitations and the kink-antikink interactions when a normal mode of a soliton gradually turns into a QNM. 
Our model approximates a large class of stability potentials allowing us to understand the main physical effects common to many solitonic models.

\begin{figure}[tbp]
\centering
 \includegraphics[width=0.84\columnwidth]{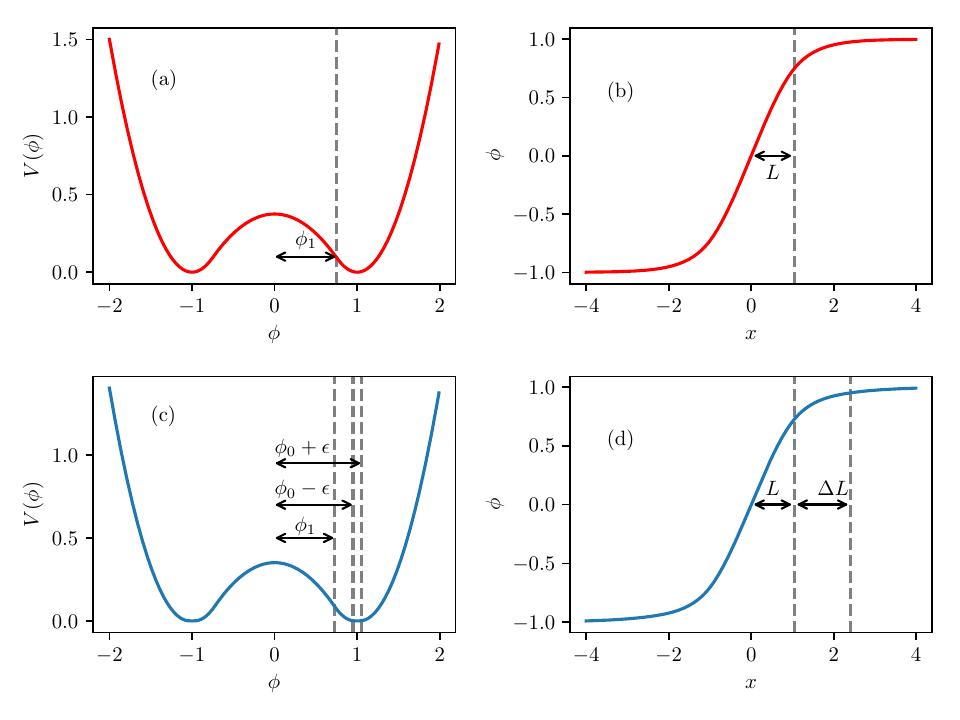}
        \caption{(a) Potential $V(\phi)$ and (b) the kink solution with $\Delta\gamma=0.0$. (c) and (d) Same as before with $\Delta\gamma=2.0$ . Dashed lines (in grey) mark the limiting points in the definition of $V(\phi)$ and $\phi(x)$. Other parameters are: (a) and (b) $\gamma=3.0(\lambda)$. (c) and (d) $\gamma=3.0(\lambda)$ and $\epsilon=0.05$.}
       \label{potential_and_kink}
\end{figure}

Now, let us design the potential such that the kink solution follows the stability equation with $U(x)$ in eq.~(\ref{stab-potential1}). We consider the following general potential 
\begin{equation} V(\phi) = 
   \begin{cases} 
      \frac{\lambda}{2}(-\phi^2+A^2), & 0\leq\phi<\phi_1, \\
      \frac{\gamma}{2}(\phi-\phi_0)^2, & \phi>\phi_1,
   \end{cases}
\label{potential1}
\end{equation}
for positive $\phi$ where all quantities are positive constants, except for $\phi$. As in the $\phi^4$ model, we impose the symmetry $V(-\phi)=V(\phi)$, which trivially extends the definition for negative $\phi$ and creates two symmetric minima at $\phi=\pm\phi_0$.  We choose the constants such that $V(\phi)$ and $dV/d\phi$ are continuous at $\phi=\phi_1$. These conditions are necessary for the field $\phi(x,t)$ and the kink solution $\phi_K(x)$ to be sufficiently well behaved. Besides that, we set $\phi_0=1$ which is equivalent to rescaling the fields $\phi$ and the constants.
In the definition of the potential in eq.~(\ref{potential1}) we have five constants but only two are independent due to the constraints imposed. 
The potential $V(\phi)$ is shown in Fig.~\ref{potential_and_kink}(a). 

The static kink solutions can be computed using the BPS condition
\begin{equation}
\frac{d\phi_K(x)}{dx}=\sqrt{2V(\phi)}.
\end{equation}
From the above condition one can see that the first and second derivatives of the kink solution are continuous. We find that the kink centered at the origin is given by
\begin{equation}
\phi_K(x)= 
  \begin{cases} 
     A\sin(\sqrt{\lambda}x), & 0<x<L, \\
     1-(1-\phi_1)e^{-\sqrt{\gamma}(x-L)}, & x>L,
  \end{cases}
\end{equation}
for positive $x$. The parameter $L$ is the position where $\phi_K(x=L)=\phi_1$, given implicitly by
\begin{equation}
\label{phi1}
\phi_1=A\sin(\sqrt{\lambda}L).
\end{equation}
Notice that $L$ is not independent of the rest of the parameters and as a result we still have only two independent parameters. A typical kink configuration in our model is shown in Fig.~\ref{potential_and_kink}(b).

We can solve eq.~(\ref{schrodinger}) resorting to the analogy with the Schr\"{o}dinger equation. The linearized potential is the well-known square well in quantum mechanics \cite{griffiths2018introduction}. Its eigenvalues for even and odd eigenfunctions are given by the following transcendental equations
\begin{align}
\label{normal1}
k=&p\tan(pL),\\
-k=&p\cot(pL),
\label{normal2}
\end{align}
respectively, where $k \equiv \sqrt{\gamma-\omega^2}$ and $p \equiv \sqrt{\omega^2+\lambda}$. The even and odd eigenfunctions are written in appendix \ref{ap2} for reference. To work with dimensionless parameters let us rescale them as $x\to x \, \sqrt{\lambda}$, $t\to t \, \sqrt{\lambda}$, $\gamma \to \gamma / \lambda$ and $\omega \to \omega/\sqrt{\lambda}$. When deem necessary, we write the scaling parameter $\lambda$ explicitly. With this rescaling we have control over only one parameter in the system.

There is at least one even solution to the transcendental equations (\ref{normal1}) and (\ref{normal2}). Therefore, the kink has at least one normal mode, which is the translational mode with $\omega^2=0$. Also there are no solutions with $\omega^2<0$. These conditions are guaranteed by the continuity of the potential. The other solutions can be found numerically and we choose the parameters such that the solution has at least one shape mode, that is, one mode with $\omega^2>0$.

At this point let us modify the potential $V(\phi)$ to create QNMs. We modify the potential for positive $\phi$ in the following form
\begin{equation} V(\phi) = 
   \begin{cases} 
      \frac{\lambda}{2}(-\phi^2+A^2), & 0\leq\phi<\phi_1, \\
      \frac{\gamma}{2}[(\phi-1+\delta)^2+B^2], & \phi_1<\phi<1-\epsilon, \\
      \frac{\gamma^\prime}{2}(\phi-1)^2, & 1-\epsilon<\phi<1+\epsilon,\\
      \frac{\gamma}{2}[(\phi-1-\delta)^2+B^2], & \phi>1+\epsilon,
   \end{cases}
\label{potential}
\end{equation}
where all quantities are positive constants, except for $\phi$.  The stability potential corresponding to the kink solution of the above potential follows eq.\ (\ref{stab-potential2}), the potential well with two barriers. We choose the potential to be still an even function of $\phi$, as the original one. We require continuity for $V$ and $dV/d\phi$ at $\phi=\phi_1$ and $\phi=1\pm\epsilon$. The potential $V(\phi)$ is shown in Fig.~\ref{potential_and_kink}(c).
By the continuity condition, $\delta\to 0$ and $B\to 0$ when $\epsilon\to 0$, the initial potential in eq.~(\ref{potential1}) is recovered. Therefore, the parameter $\epsilon$ encodes the difference between the two stability potentials and as we will see shortly the transition between the normal and QNMs. The minimum of the potential is set again at $\phi=1$. 
In the definition of the potential in eq.~(\ref{potential}) there are eight constants but only four are independent, because of the constraints imposed by the continuity conditions. With the same rescaling as in the original potential, there are three parameters one can control to study the system. 

The static kink solutions can be computed again. We find that the kink centered at the origin is given by
\begin{equation}
\phi_K(x)= 
  \begin{cases} 
     A\sin x, & 0<x<L, \\
     1-\delta+B\sinh(C+\sqrt{\gamma}(x-L)), & L<x<L^\prime, \\
     1-\epsilon e^{-\sqrt{\gamma^\prime}[x-(L+\Delta L)]}, & x>L^\prime,
  \end{cases}
\end{equation}
for positive $x$, where 
\begin{equation}
C\equiv\ln(B/D),
\end{equation}
and
\begin{equation}
D\equiv\sqrt{B^2+(1-\delta-\phi_1)^2}+1-\delta-\phi_1.
\end{equation}
$L$ and $L^\prime\equiv L+\Delta L$ are the points where $\phi_K=\phi_1$ and $\phi_K=1-\epsilon$, respectively, and are not independent of the rest of the parameters. A typical kink configuration is shown in Fig.~\ref{potential_and_kink}(d).

\section{Stability Equation}
\label{stability}

We plan to study the stability equation (\ref{schrodinger}) for the general case where $\epsilon\neq 0$. Let us look at scattering solutions with $\gamma-\Delta\gamma<\omega^2<\gamma$. We can write the solutions in the different regions of space as
\begin{equation}
\label{coef}
\eta(x)= 
  \begin{cases} 
     F_1e^{imx}+F_2e^{-imx}, & x<-L-\Delta L \\
     G_1e^{-kx}+G_2e^{kx}, & -L-\Delta L<x<-L \\
     H_1\sin(px)+H_2\cos{px}, & -L<x<L \\
     I_1e^{-kx}+I_2e^{kx}, & L<x<L+\Delta L \\
     Je^{imx}, & x>L+\Delta L
  \end{cases}
\end{equation}
where $m \equiv \sqrt{\omega^2+\Delta\gamma-\gamma}$ and the capital letters $F$-$J$ are the constants that we need to find. As we mentioned before, $k \equiv \sqrt{\gamma-\omega^2}$ and $p \equiv \sqrt{\omega^2+\lambda}$. Boundary conditions are given by requiring continuity of $\eta(x)$ as well as its first derivative at $x=\pm L$ and $x=\pm (L+\Delta L)$. The constants $F$-$J$ are presented in appendix \ref{ap1}. After some algebra, one can find the transmission and reflection coefficients
\begin{align}
\label{transmission}
\begin{split}
T=& 4\bigg\rvert e^{2k\Delta L}(1+i\alpha^-_{km})[\cos(2pL)+\alpha^-_{kp}\sin(2pL)]-2i\alpha^+_{km}\alpha_{kp}^+\sin(2pL)\\
  &+e^{-2k\Delta L}(1-i\alpha_{km}^-)[\cos(2pL)-\alpha_{kp}^-\sin(2pL)]\bigg\rvert^{-2},
\end{split}\\
\begin{split}
R=& \frac{T}{4}\bigg\{e^{2k\Delta L}\alpha_{km}^+[\cos(2pL)+\alpha_{kp}^-\sin(2pL)]-2\alpha_{km}^-\alpha_{kp}^+\sin(2pL)\\
  &-e^{-2k\Delta L}\alpha_{km}^+[\cos(2pL)-\alpha_{kp}^-\sin(2pL)]\bigg\}^{2},
\end{split}
\end{align}
where
\begin{align}
\label{defin}
\alpha_{km}^\pm\equiv\frac{k^2\pm m^2}{2mk},\quad
\alpha_{kp}^\pm\equiv\frac{k^2\pm p^2}{2pk}.
\end{align}

From these analytical results it is easy to find the QNMs. They are defined as states with purely outgoing boundary conditions, that is, they are given by eq.~\ref{coef} with $F_1=0$. This coefficient is given in appendix \ref{ap1}. Setting it equal to zero results in
\begin{equation}
\label{QNM}
\begin{split}
&e^{2k\Delta L}[1+i\alpha_{km}^-][\cos(2pL)+\alpha_{kp}^-\sin(2pL)]-2i\alpha_{km}^+\alpha_{kp}^+\sin(2pL)\\
&+e^{-2k\Delta L}[1-i\alpha_{km}^-][\cos(2pL)-\alpha_{kp}^-\sin(2pL)]=0,
\end{split}
\end{equation}
which gives the quasinormal frequencies implicitly. The real part of the frequency, $\omega$, corresponds to the oscillation frequency $\Omega$ and the imaginary part to the decay rate $\Gamma$. Similarly, we can compute the bound state frequencies by setting $im\to-q$, where $q \equiv \sqrt{\gamma-\Delta\gamma-\omega^2}$, as well as $F_1=0$ in eq.~(\ref{coef}). This is the same as setting $im\to-q$ in eq.~(\ref{QNM}).

\begin{figure}[tbp]
\centering
   \includegraphics[width=0.625\columnwidth]{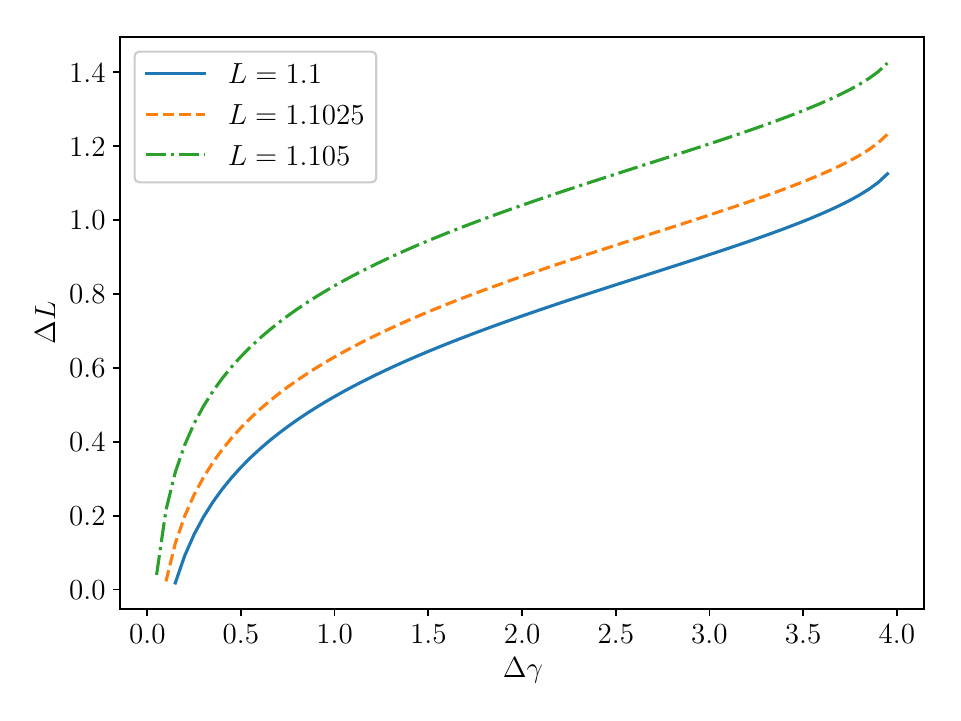}
        \caption{Relationship between parameters of the linearized potential, $\Delta\gamma$ versus $\Delta L$, for different values of $L$ considering $\gamma=4.0$.}
       \label{DGDL}
\end{figure}

We plan to investigate how the transmission and reflection coefficients $T$ and $R$, as well as the bound and quasinormal frequencies, depend on the parameters of our model. Ideally, one could fix all the parameters of the linearized potential ($\gamma$, $\Delta\gamma$, $L$, $\Delta L$) but one, to isolate the effect of each variable. However, this is not possible due to the fact that after rescaling there are only three independent variables. 
Hence, we fix $L$ and $\gamma$ and isolate the effect of $\Delta\gamma$ and $\Delta L$. These two parameters are not independent from each other and their relation can by computed numerically from the continuity conditions. $\Delta L$ as a function of $\Delta\gamma$ is shown in Fig.~\ref{DGDL} considering $\gamma=4.0$ and three values of $L=1.1$, $1.1025$, $1.105$.

\begin{figure}[tbp]
\centering
    \includegraphics[width=1.0\columnwidth]{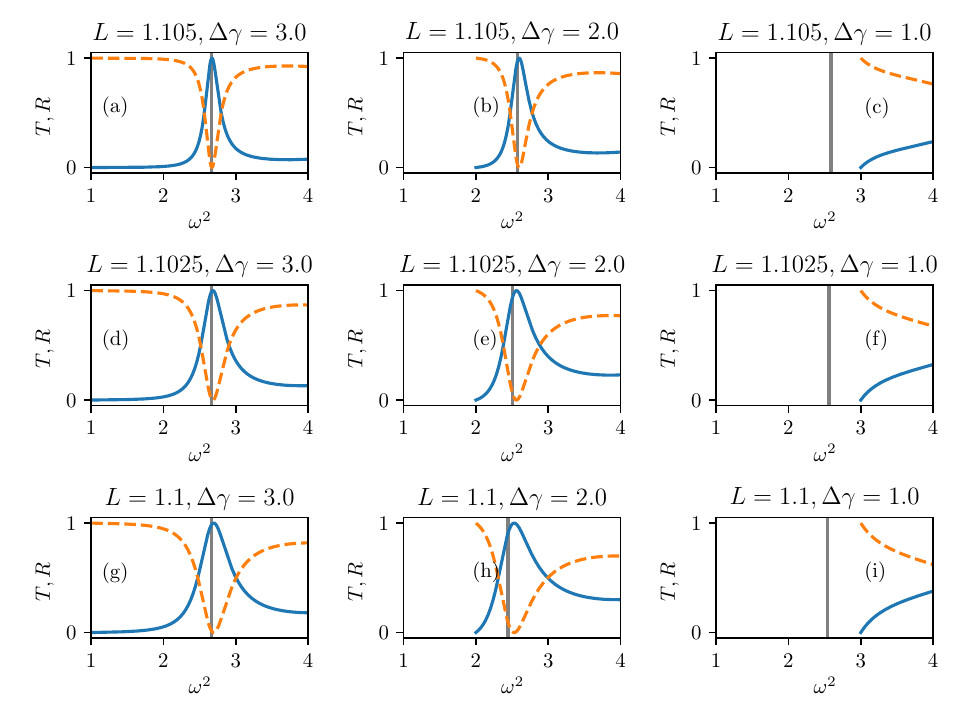}
        \caption{Transmission (solid) and reflection (dashed) coefficients as a function of the incident frequency for an incident wave in a barrier with a square well in the center. We vary $\Delta\gamma$ and $L$ and set $\gamma=4.0$. Vertical grey lines are the QNM or bound state frequencies}. The peaks are not exactly centered at the grey lines, but becomes more localized around it as $L$ increases.
       \label{transm}
\end{figure}

The transmission and reflection coefficients $T$ and $R$ are shown in Fig.~\ref{transm} for several values of the parameters. We find numerically that, when $\omega$ is close to, but not exactly at, the quasinormal oscillation frequency $\Omega$, there appears a resonance where the barrier becomes almost transparent, i.e. $T\approx1$, $R\approx0$, which is an interesting result. The quasinormal oscillation frequencies are marked as vertical lines in Figs.~\ref{transm}(a), (b), (d), (e), (g) and (h). The peak in the transmission coefficient becomes slightly more localized around the square well normal modes as $L$ and/or $\Delta \gamma$ (or equivalently $\Delta L$) increase. Physically, one can understand this behavior by noting that only if the incident frequency is near the QNM frequency it could excite the system into this state, due to resonance, and tunnel through the barrier. Moreover, for higher values of $\Delta L$ we need a more precise tuning of the incident frequency to excite the system into this state. The conclusion is that the existence of QNMs are linked to resonance peaks, as expected.

For small values of $\Delta\gamma$, see for example Figs.~\ref{transm}(c), (f) and (i), there is neither tunneling nor resonance because there is no QNM in the scattering frequencies range. In fact there is no QNM at all for small values of $\Delta\gamma$. There is a bound state instead which is marked as vertical lines in the graphs.

\begin{figure}[tbp]
\centering
    \includegraphics[width=0.8\columnwidth]{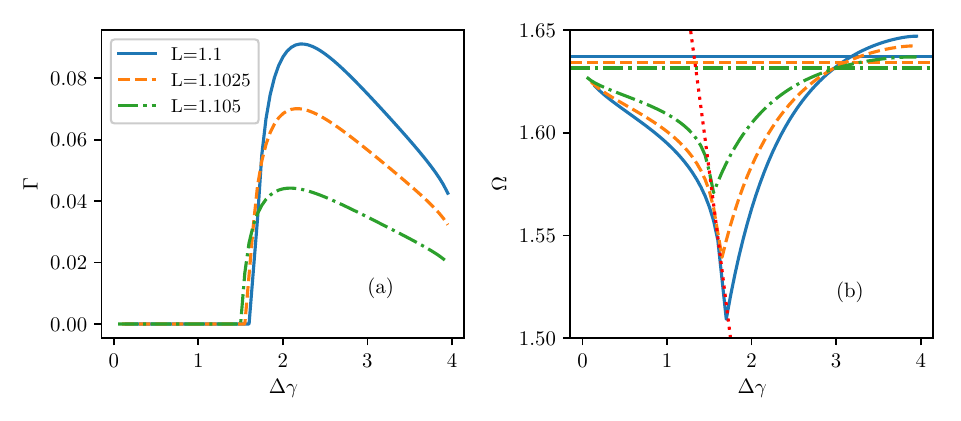}
    \caption{(a) and (b) Analytical value of exponential decay rate and the frequency of oscillation, respectively, as a function of $\Delta\gamma$. Parameters are $L=1.1,1.1025,1.105$ and $\gamma=4.0$ in both graphs.}
       \label{decay}
\end{figure}

At this point let us focus on computing the oscillation frequencies $\Omega$ and the decay rates $\Gamma$ considering $F_1=0$ as we mentioned before. Trivially, if one computes the bound states frequencies, there always appears a zero mode with $\omega=0$. For small values of $\Delta\gamma$ the linearized potential is very close to the square well. As before we adjust the parameters of the system to have an excited bound state in the square well limit. In Fig.~\ref{decay}(a) we plot the decay rate as a function of $\Delta\gamma$. We see that the decay rate is zero for small values of $\Delta\gamma$ which is the case close to the square well. Increasing $\Delta\gamma$, at some point the excited bound state turns into a QNM leading to $\Gamma > 0$ and creating a discontinuity in the derivative. After the initial increase the decay rate $\Gamma$ peaks at some value and then starts to decrease. This occurs because the $\Delta L$ is increasing (see Fig.~\ref{DGDL}) which means that the QNM has to tunnel a larger barrier.

In Fig.~\ref{decay}(b) we see the frequency of oscillation $\Omega$ versus $\Delta\gamma$. We also show for comparison the frequencies $\omega_1=1.637,1.634,1.632$ of the first excited state of the square well with $L=1.1,1.1025,1.105$, respectively, and $\gamma=4.0$. For small values of $\Delta\gamma$, there are only bound state frequencies. These frequencies start close to the frequencies of the square well $\omega_1$. As $\Delta\gamma$ increases $\Omega$ decreases because the potential energy is decreasing. This happens until the bound state disappears and the QNM appears at the point with discontinuous derivative. After that $\Omega$ increases because $\Delta L$ is increasing and the potential becomes closer to the square well again. There exists a frequency $\omega$ that solves both the QNM equation (\ref{QNM}) and the bound state equation. If this is true, then $im=-q$. Therefore, we have that $m=q=0$ or $\omega=\sqrt{\gamma-\Delta\gamma}$. This is the dotted curve shown in Fig.~\ref{decay}(b).  Some values of $\Gamma$ and $\Omega$ for different parameters of the model are given in Table \ref{QNMfreq} for future reference.

We find similar results for $\Gamma(\Delta\gamma)$ and $\Omega(\Delta\gamma)$ integrating the equations of motions or linearized equations of motion numerically with square well bound states as initial conditions. However, close to the point of the transition from the bound frequencies to QNM ones the results become highly inaccurate compared with the analytical ones discussed before. When we integrate numerically we see that the discontinuities in the derivatives of $\Gamma$ and $\Omega$ are smoothed out, making the result untrustworthy.

\begin{table}
\centering
\begin{tabular}{c|c|c|c|c|l}
$\gamma$ & $\epsilon$ & $L$ & $\Delta\gamma$ & $\Gamma$ & $\Omega$\\
\hline
4.0 & & 1.1 & 1.0 & 0.0 & 1.595\\
4.0 & & 1.1 & 2.0 & 0.087 & 1.563\\
4.0 & & 1.1 & 3.0 & 0.075 & 1.632 \\
4.0 & & 1.1025 & 1.0 & 0.0 & 1.601\\
4.0 & & 1.1025 & 2.0 & 0.068 & 1.584\\
4.0 & & 1.1025 & 3.0 & 0.057 & 1.632 \\
4.0 & & 1.105 & 1.0 & 0.0 & 1.609\\
4.0 & & 1.105 & 2.0 & 0.044 & 1.605 \\
4.0 & & 1.105 & 3.0 & 0.035 & 1.631\\
2.667 & 0.0 & & & 0.0 & 1.533 \\
2.667 & 0.005 & & 2.0 & 0.0067 & 1.538 \\
2.667 & 0.02 & & 2.0 & 0.021 & 1.547 \\
2.667 & 0.04 & & 2.0 & 0.040 & 1.557 \\
2.667 & 0.06 & & 2.0 & 0.059 & 1.566 
\end{tabular}
\caption{Values of the oscillation frequency $\Omega$ and decay rate $\Gamma$ for the different values of the parameters used in the figures.}
\label{QNMfreq}
\end{table}

\section{Collision}
\label{collision}

Let us now study the collision of the kink and antikink in our model, considering both cases with normal and QNMs. In order to do that, we need to integrate the field equations numerically to find the field $\phi(x,t)$. For details on the numerical implementation refer to Appendix \ref{ap3}. We set the following additive ansatz as the initial condition of the field 
\begin{align}
\phi(x,0)&=\phi_{\bar{K}}\bigg(\frac{x}{\sqrt{1-v_i^2}}-X_0\bigg)+\phi_{K}\bigg(\frac{x}{\sqrt{1-v_i^2}}+X_0\bigg)-1\\
\dot{\phi}(x,0)&=\frac{v_i}{\sqrt{1-v_i^2}}\bigg[\phi^\prime_{\bar{K}}\bigg(\frac{x}{\sqrt{1-v_i^2}}-X_0\bigg)-\phi^\prime_{K}\bigg(\frac{x}{\sqrt{1-v_i^2}}+X_0\bigg)\bigg],
\end{align}
where $\phi_{\bar{K}}(x)$ is the antikink static solution and prime and dot denote derivatives with respect to the position and time, respectively. We vary the initial velocity of the kink and antikink $v_i$ and set the initial position at $X_0=10.0$. 

As there are three independent parameters, we choose the simplest way to create QNMs, fixing $\gamma$ and $\Delta\gamma$ and varying $\epsilon$. We take a large enough $\Delta\gamma$ to guarantee the existence of QNMs for $\epsilon\neq 0$. In the following, first we consider the collision for $\epsilon=0$ and show that it presents resonance windows that are well described by the resonant energy exchange mechanism of Campbell et al. \cite{campbell1983resonance}. Then, we consider the collision for $\epsilon\neq 0$, where normal modes become QNMs.

\subsection{Potential with $\epsilon=0$}
\label{zero}

\begin{figure}[tbp]
\centering
   \begin{subfigure}[b]{0.32\linewidth}
     \includegraphics[width=\linewidth]{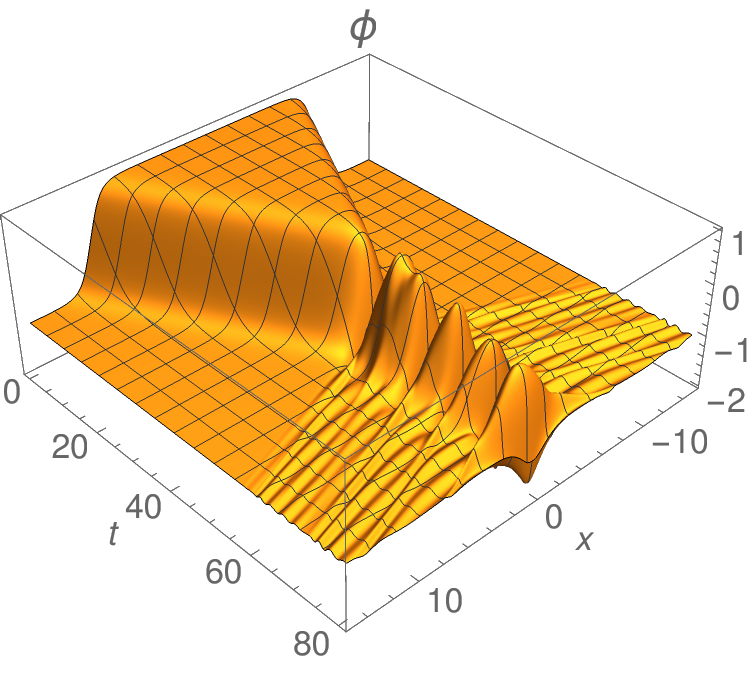}
     \caption{}
   \end{subfigure}
   \begin{subfigure}[b]{0.32\linewidth}
     \includegraphics[width=\linewidth]{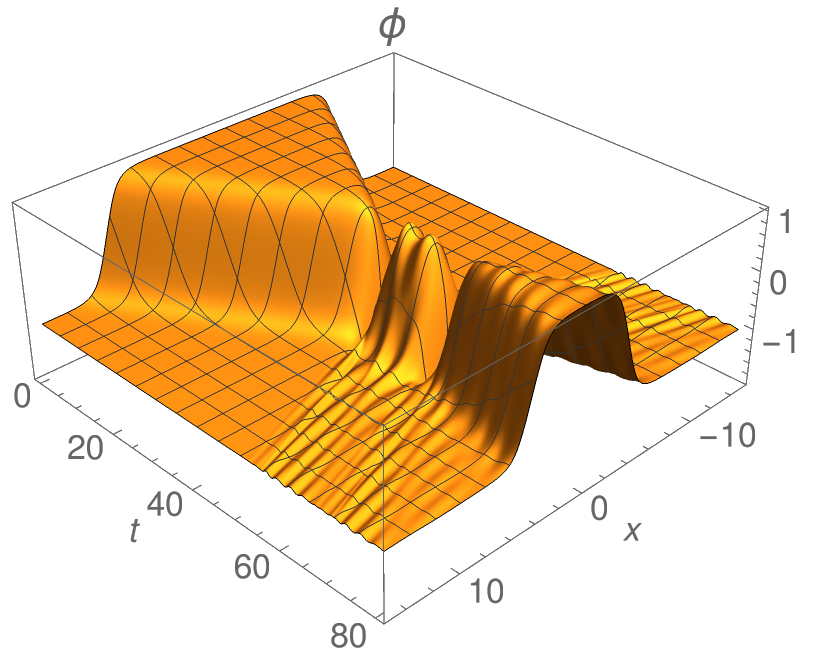}
     \caption{}
   \end{subfigure}
   \begin{subfigure}[b]{0.32\linewidth}
     \includegraphics[width=\linewidth]{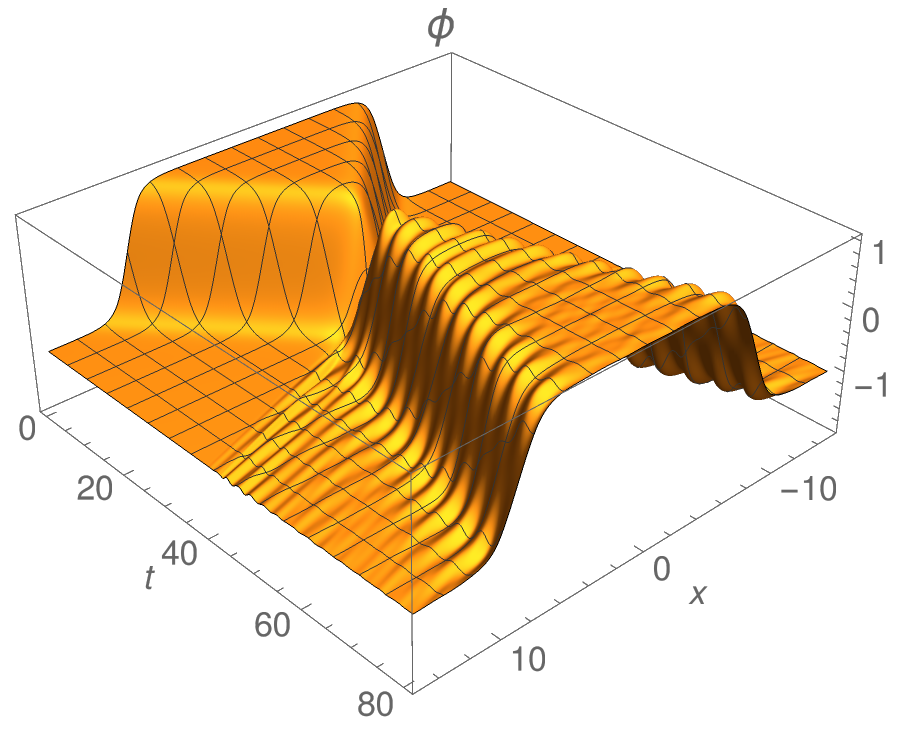}
     \caption{}
   \end{subfigure}
        \caption{Evolution of the scalar field $\phi(x,t)$ during the collision of a kink and an antikink. Parameters are $\epsilon=0$, $\gamma=2.667$ and (a) $v_{in}=0.200$, (b) $v_{in}=0.228$ and (c) $v_{in}=0.340$. In (a) the kink and the antikink annihilate, in (b) they bounce two times before separating and in (c) they reflect.}
       \label{draw}
\end{figure}

\begin{figure}[tbp]
\centering
  \includegraphics[width=\columnwidth]{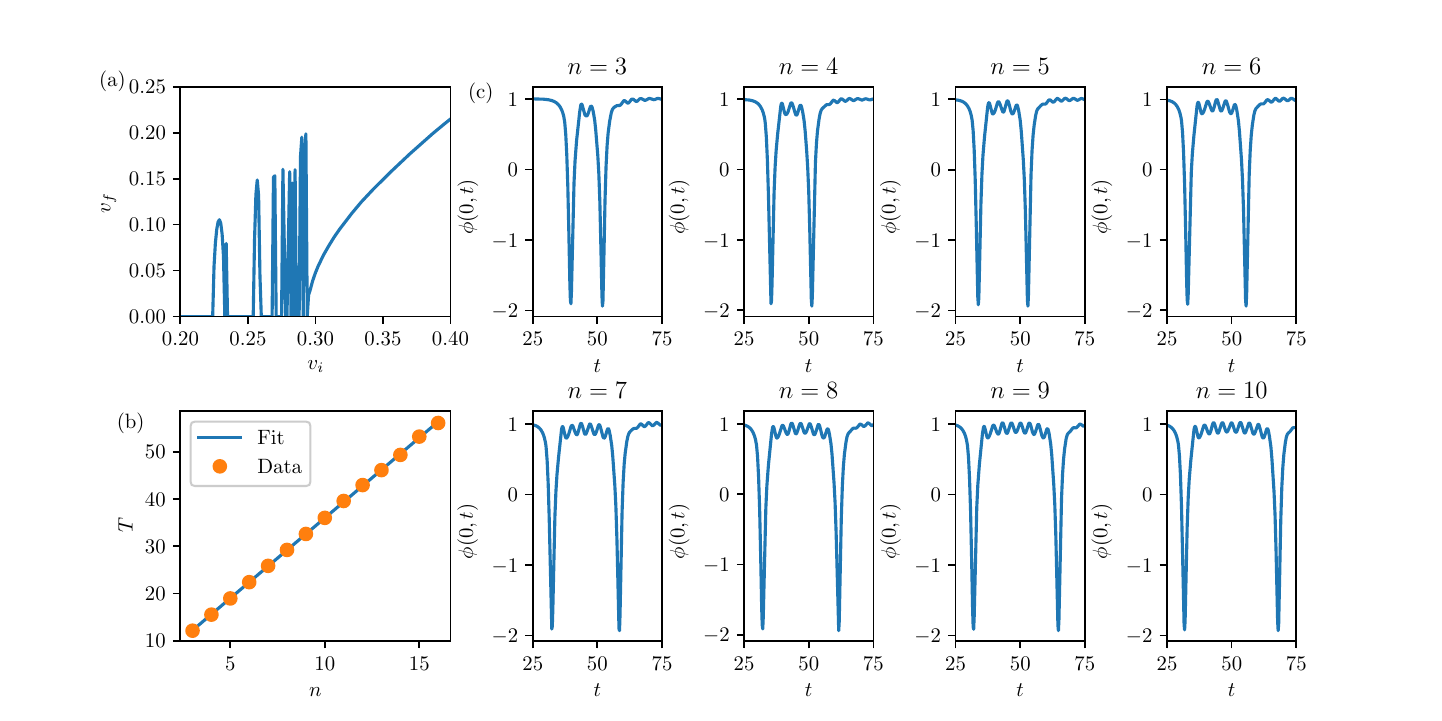}
        \caption{ (a) Final velocity after the collision versus the initial velocity of the kink and antikink. (b) Evolution of the center of the scalar field $\phi(0,t)$ during the collision of a kink and an antikink for several bounce windows where $n$ is the window number. (c) Plot of the time between the first and second collisions versus the window number $n$. Parameters in all graphs are $\epsilon=0$ and $\gamma=2.667$.}
       \label{T}
\end{figure}

Setting $\epsilon=0$ gives the case of collisions of a kink with a translational and vibrational normal modes which are solutions of the square well potential. We choose the parameters such that the kink has only one normal mode. In the collision of a kink with an antikink in the system there are three possible behaviors: the kink and antikink may annihilate, resonate or reflect. It is worth mentioning that our model has well behaved collisions that exhibit the expected behavior of a kink and antikink collision for a potential similar to the one in $\phi^4$ theory. In Fig.~\ref{draw}(a) the kink and antikink collide and then turn into a bion that oscillates and decays slowly through radiation. In Fig.~\ref{draw}(b) the kink and antikink bounce two times before separating in opposite directions. Separation after multiple bounces is called resonance. Finally, in Fig.~\ref{draw}(c) the kink and  antikink collide once and then reflect.
Resonances occur in resonance windows for velocities below a critical velocity, above which the kink and antikink reflect. These windows alternate between intervals where the kink and antikink annihilate. This behavior is clear when we look at the plot of the final velocity of the kink and antikink $v_f$ versus the initial velocity $v_i$ shown in Fig.~\ref{T}(a). The final velocity of the kink is computed by measuring its positions after crossing $x=15.0$, during an interval $\Delta t=20.0$ with spacing $\delta t=0.2$. Finally, we fit the points into a straight line where the slope of the line is the kink velocity. This method minimizes the error due to fluctuations in the kink velocity. The errors in the slope are quite small, of the order $10^{-4}$. When $v_f=0$, it means that the kink and antikink annihilated. The sharp peaks with $v_f\neq0$ are the resonance windows. As we mentioned before, there is a critical velocity $v_c$ above which the kink and antikink always reflect. Taking $\epsilon=0$ and $\gamma=2.667$, the critical velocity has the value $v_c=0.298$.

The resonance phenomenon obeys the resonant energy exchange mechanism of Campbell et al. \cite{campbell1983resonance} as shown in Figs.~\ref{T}(b) and (c). In short, according to the mechanism after the first collision part of the translational energy is temporarily transferred to vibrational modes and the kink and antikink may be forced to collide once more.
Since the vibrational mode is localized at the kink,
if the timing is right the second collision transfers the energy back to the translational mode and the kinks fully separate.

In Fig.~\ref{T}(b) one can see that the time interval between the two bounces $T$ increases by one oscillation of the field at the center of the collision $\phi(0,t)$ as we go from one resonance window to the next. As a result, $T$ must fall in a straight line when plotted as a function of the window number $n$ as shown in Fig.~\ref{T}(c). One can write this relation in the form
\begin{equation}
\omega_1 T=\delta+2\pi n,
\end{equation}
where $\omega_1$ is the oscillation frequency of the first excited state of the square well potential and $\delta$ is a phase \cite{campbell1983resonance}. The window number of the first two-bounce window is chosen such that the phase $\delta$ is between $0$ and $2\pi$ and it increases by one from one window to the next. Fitting the curve, we find $\delta=3.719$ and $\omega_1=1.510$ which is close to the theoretical value $\omega_1=1.533$ obtained by solving the stability equation (\ref{schrodinger}) for the same values of the parameters. 

\subsection{Potential with $\epsilon\neq 0$}
\label{notzero}

\begin{figure}[tbp]
\centering
  \includegraphics[width=0.7\columnwidth]{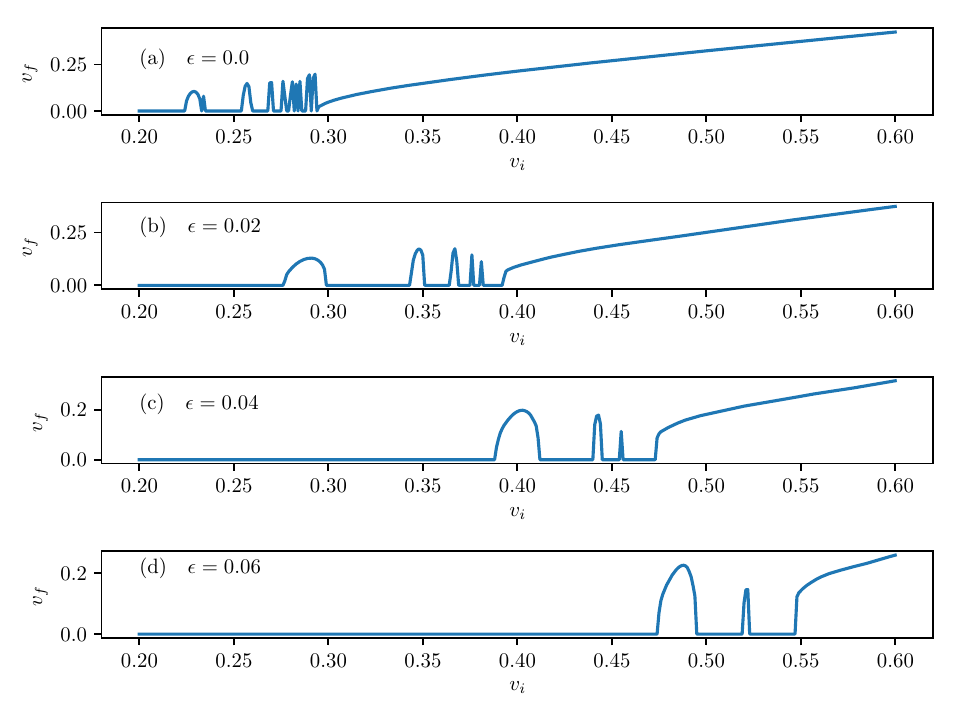}
        \caption{Final velocity of the kink and antikink $v_f$ as a function of the initial velocity $v_i$ for: (a) $\epsilon=0$, (b) $\epsilon=0.02$, (c) $\epsilon=0.04$ and (d) $\epsilon=0.06$. As $\epsilon$ increases the critical velocity also increases and higher order resonance windows are suppressed. In all graphs we consider $\gamma=2.667$, $\Delta\gamma=2.0$.}
       \label{vinvout}
\end{figure}

Now we study kink-antikink collisions for potentials with $\epsilon\neq 0$. This means that the vibrational modes become QNMs and thus will tunnel the potential barrier created by the kink. The final velocity of the kink and antikink $v_f$ as a function of the initial velocity $v_i$ for several values of $\epsilon$ is shown in Fig.~\ref{vinvout}. 
We also find that the critical velocity increases as $\epsilon$ increases from zero, consistent with the result found in \cite{dorey2018resonant}. For example, the critical velocities are $v_c=0.394,0.474,0.548$ for $\epsilon=0.02,0.04,0.06$, respectively. It means that,  for higher values of $\epsilon$, the kink and antikink need higher relative initial velocities to be able to reflect. Moreover, most resonance windows are suppressed as the normal modes become QNMs and one finds a much poorer structure. 
This happens because for $\epsilon\neq 0$ translational energy is leaked when it is converted into vibrational energy and thus it cannot be recovered. 

\begin{figure}[tbp]
\centering
    \includegraphics[width=0.8\columnwidth]{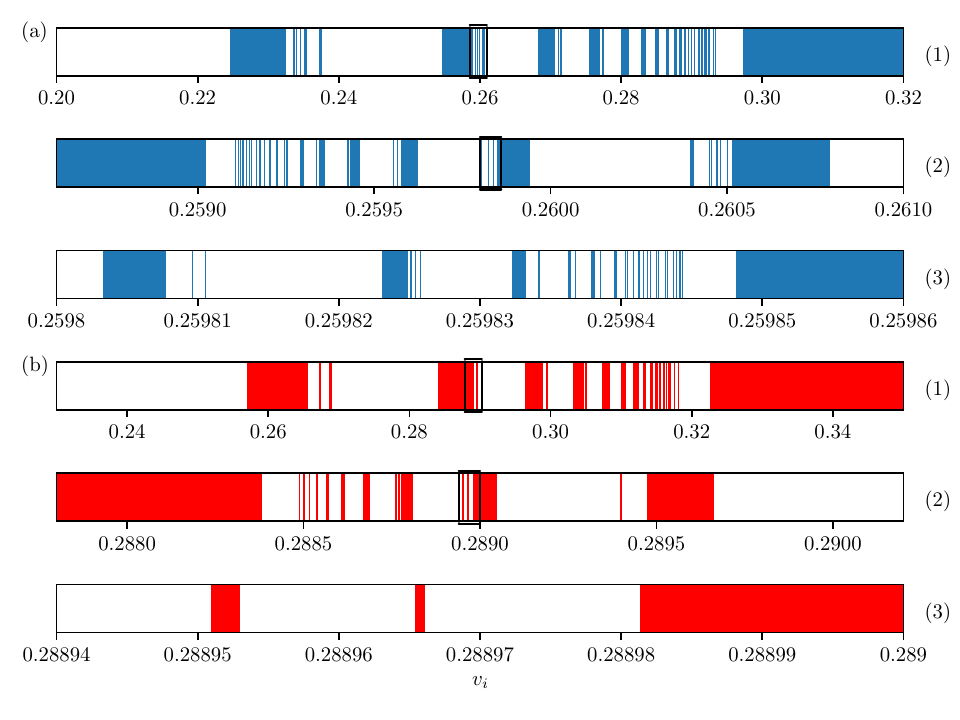}
        \caption{Resonance windows of the collision between a kink and an antikink for (a) $\epsilon=0.0$ and (b) $\epsilon=0.005$. As we zoom in, the windows appear to be self-similar, except for the second zoom in (b). In all graphs we consider $\gamma=2.667$, $\Delta\gamma=2.0$.}
       \label{fractal}
\end{figure}

We can also investigate the self-similarity of the resonance windows in our model. By self-similar, we mean that the structure of the resonance windows is similar across different scales.
In Fig.~\ref{fractal} we show the resonance windows intervals for $\epsilon=0.0,0.005$. Fig.~\ref{fractal}(a), corresponding to $\epsilon=0.0$, exhibits the self-similarity in the model. As we zoom in near resonance windows, we observe the same structure of higher-bounce resonance windows analogous to the one found in the $\phi^4$ model \cite{anninos1991fractal}. 
Near two-bounce resonance windows we find a similar structure of three-bounce resonance windows, near three-bounce resonance windows we find a similar structure of four-bounce resonance windows and so on. 
This situation contrasts with the one depicted in Fig.~\ref{fractal}(b), for  $\epsilon=0.005$. Although after the first zoom the resonance windows still appear to be self-similar, we see that this is not true for a second zoom. We barely observe any four-bounce resonance windows.
As we increase $\epsilon$, we lose the self-similarity of three-bounce resonance windows. Increasing even more, most of the two-bounce resonance windows are suppressed.
In fact, there is a gradual loss of similarity as $\epsilon$ increases from zero. 
This is due to the fact that for larger $\epsilon$ more energy is leaked, leaving less energy available for the next bounce. As a result, multiple bounce becomes less probable, breaking the self-similarity.

\begin{figure}[tbp]
\centering
   \begin{subfigure}[b]{0.49\linewidth}
     \includegraphics[width=\linewidth]{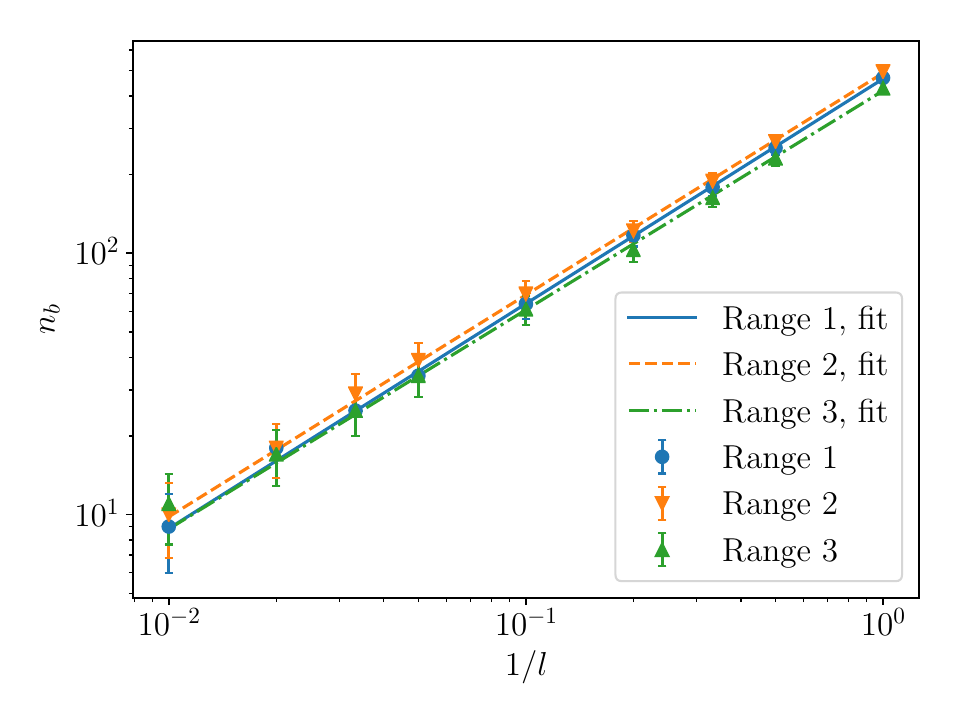}
     \caption{$\epsilon=0.0$}
   \end{subfigure}
   \begin{subfigure}[b]{0.49\linewidth}
     \includegraphics[width=\linewidth]{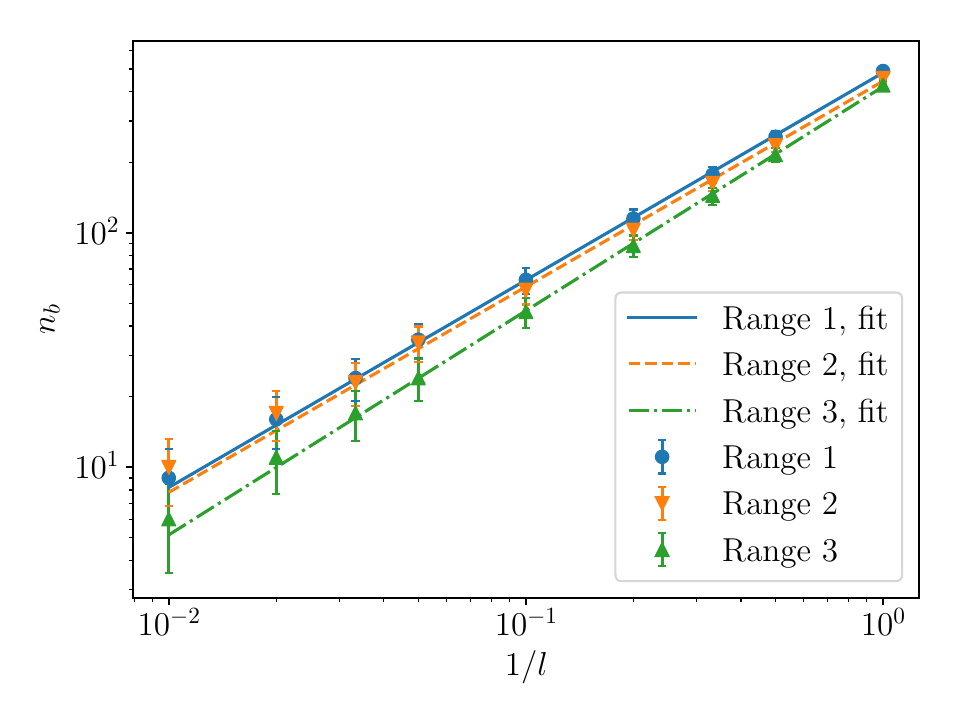}
     \caption{$\epsilon=0.005$}
   \end{subfigure}
        \caption{Number of boxes to cover the set of resonance windows intervals shown in Fig.~\ref{fractal} versus the inverse of the size of the boxes in log scale. The box lengths are rescaled to lie in the same range. Parameters are $\gamma=2.667$ and $\Delta\gamma=2.0$.}
       \label{boxes}
\end{figure}

The fractal self-similarity can be quantified in two ways. First, we compute the fractal dimension of the intervals shown in Fig.~\ref{fractal}. We divide the interval in boxes and we count the number of boxes $n_b$ needed to cover the resonance windows for different box lengths $l$. The fractal dimension is given by $D=-\lim_{l\to 0}\log{n_b}/\log{l}$. The plot of $n_b$ versus $1/l$ is shown in log scale in Fig.~\ref{boxes} where the fractal dimension is estimated by the slope of the curve. The values of the slopes are shown in table \ref{dimension}.
 We see, considering $\epsilon=0.0$, that all three intervals shown in Fig.~\ref{fractal} (a) have almost the same slope according to this analysis. This means that in this case one can associate fractal structure to the multibounce resonance windows with fractal dimension approximately in the range $0.84-0.86$. The fractal dimension of this set of intervals is an evidence that the structure is self-similar. Of course there are finite size effects as we divided the interval in a finite number of boxes and the true fractal dimension is only found in the limit when the number of boxes goes to infinity.
 In contrast, considering $\epsilon=0.005$, the slope is slightly larger than when $\epsilon=0.0$ in the two largest intervals, range (1) and (2) in Fig.~\ref{fractal} (b).  Besides that, it does not keep the same slope in the three intervals. As we zoom in to the smallest interval the slope is approximating to unity. This means that the $\epsilon=0.005$ case fails to exhibit self-similarity at some point. As a matter of fact, this is true in general for any $\epsilon \neq 0$.

\begin{table}
\centering
\begin{tabular}{c|c|c|c|l}
Range number&$\epsilon$ & Range in $v_i$ & Smallest box size & Slope\\
\hline
1 &$0.0$ & $0.20-0.32$ & $1\times 10^{-4}$ & $0.860\pm 0.006$\\
2 &$0.0$ & $0.2586-0.2610$ & $2\times 10^{-6}$ & $0.849\pm 0.006$\\
3 &$0.0$ & $0.25980-0.25986$ & $5\times 10^{-8}$ & $0.84\pm 0.01$\\
\hline
1 &$0.005$ & $0.23-0.35$ & $1\times 10^{-4}$ & $0.885\pm 0.008$\\
2 &$0.005$ & $0.2878-0.2902$ & $2\times 10^{-6}$ & $0.88\pm 0.02$\\
3 &$0.005$ & $0.28894-0.28900$ & $5\times 10^{-8}$ & $0.956\pm 0.009$
\end{tabular}
\caption{Slope of $n_b$ as a function of $1/l$ (inverse of the box length) in log scale for three different ranges in $v_i$ for two values $\epsilon=0.0$ and $\epsilon=0.005$. Each range is divided up to maximum 1200 boxes which gives the smallest box size in the table. The range numbers in the table correspond to the labels in Fig.~\ref{fractal}. Parameters are $\gamma=2.667$ and $\Delta\gamma=2.0$.}
\label{dimension}
\end{table}

\begin{figure}[tbp]
\centering
    \includegraphics[width=0.6\columnwidth]{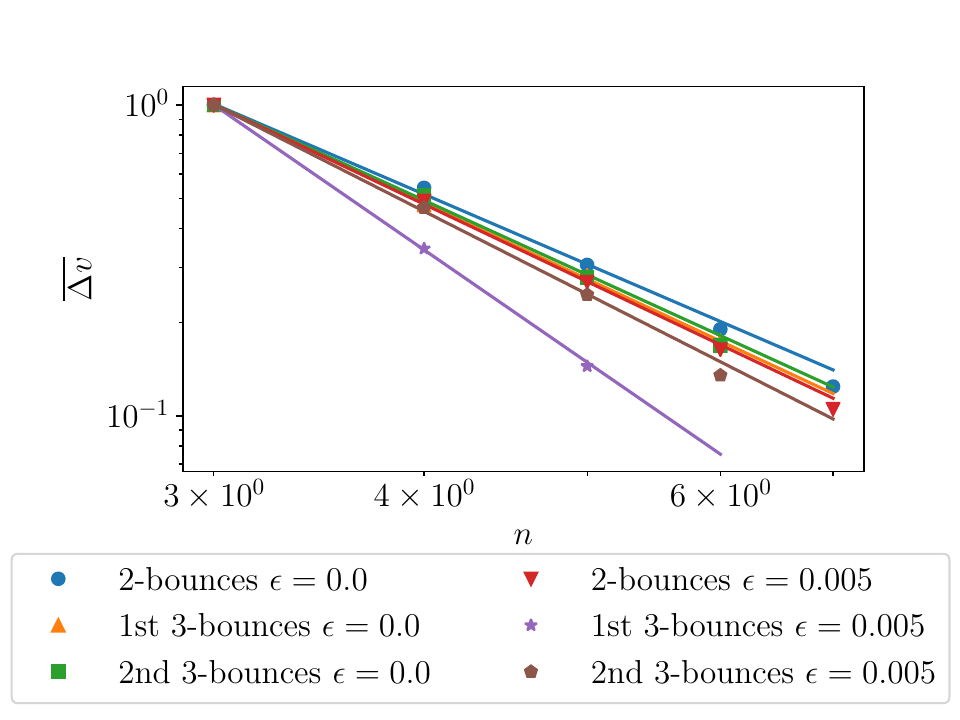}
        \caption{Normalized width of the resonance windows versus window number in log scale for the two-bounce windows and some three-bounce windows. The figure shows approximately the same slope for $\epsilon=0.0$, in contrast with $\epsilon=0.005$ case.
Parameters are $\gamma=2.667$, $\Delta\gamma=2.0$ and $\epsilon=0.0,0.005$.}
       \label{colapse}
\end{figure}

The second quantitave measure of the self-similarity of the resonance windows is given by the graph of the resonance windows width $\overline{\Delta v}$ as a function of the window number $n$. This is shown in Fig.~\ref{colapse}. The window width $\overline{\Delta v}$ is normalized by the size of the window with $n=3$. The window number of the higher-bounce windows is given by the number of field oscillations between the last two collisions. In \cite{anninos1991fractal}, the authors were able to show that the window widths obey a scaling relation, which is the same for higher-bounce resonance windows. Here, we find a scaling relation only approximately, as indicated by the fits. Moreover, we see in the figure that, when $\epsilon=0.0$, the dependence of $\overline{\Delta v}$ in $n$ is approximately the same for three-bounce resonance windows and the two-bounce resonance windows. This suggests that the resonance intervals are self-similar. However, for $\epsilon=0.005$, $\overline{\Delta v}$ decays faster for higher-bounce windows, specially the first three-bounce resonance windows, that is, the three-bounce resonance windows near the first two-bounce resonance window. This means that self-similarity is lost and occurs because this set of windows were suppressed by the leakage of vibrational energy when the QNMs are excited.

We conclude that when the normal modes of the kink become QNMs the resonance windows are suppressed. This occurs because, as is well known, the normal modes are very important for this effect. As the kink and the antikink collide, some of the translational energy is converted into vibrational energy which, if there is no leakage, keeps localized at the kink. Thus, this energy can be recovered at consecutive collisions. However, if we turn the normal mode of the kink into a QNM, part of the vibrational energy of the kink is leaked as shown in section \ref{stability} and therefore less energy can be converted back to translational energy. As a result, the resonance windows are suppressed. Furthermore, this gradually destroys the self-similar structure of the resonance windows, as expected. The critical velocity is also affected by the change from normal modes to QNMs. In the first collision between the kink and the antikink more energy is lost because the vibrational energy is now leaked and they need higher initial velocity to be able to separate after that collision.

\section{Conclusion}
\label{conclusion}

In the present paper, we have designed a scalar field model in (1+1) dimensions which has a kink and an antikink interpolating between the two symmetric minima of the potential. This model has a stability equation for perturbations of the kink which is analogous to the Schr\"{o}dinger equation with a linearized potential in the shape of the square well potential in quantum mechanics. We have designed another potential which is a modification of the original one with the stability potential in the form of a square well potential with two barriers. In this case, vibrational normal modes of the kink solutions of the former potential turns into QNMs in the latter one. The analogy with the Schr\"{o}dinger equation in our model is very clear and it has been used to show analytically that the system has QNMs, which are quasibound states that decay with time and leak energy. Moreover, these modes create resonances in the transmission coefficient of the linearized potential. The idea of constructing this peculiar model from the beginning and showing that it works to study kink-antikink collisions is the proximity of the linear stability potential defined here and the ones associated to many known kink models in the literature. This makes the model a rich ground to study and understand the physics of a large class of kinks in this context for normal and QNMs.
This model is particularly interesting because it has several parameters that can be varied to study the system dependence on several potential and linearized potential shapes. However, we have not exhausted the whole parameter space of the model in this work due to the relevance to our study.

Although our potential is piecewise defined with manufactured linear stability potential making it exotic, using the continuity condition for the potential and its derivative with respect to the scalar field as well as the BPS condition, it is well behaved and captures the most relevant characteristics of the known kink solutions studied in the literature. In fact, it behaves similarly to the $\phi^4$ kink and the corresponding kink-antikink collisions considering our original model with the square well stability potential, $\epsilon=0$. The parameter $\epsilon$ encodes the information about the size of the barrier, that is, when it increases from zero the normal modes of the square well potential may turn into the QNMs due to the appearance of the barriers causing the energy leak.

In the limit $\epsilon=0$, we have shown that in a collision between a kink and an antikink the system exhibits resonance windows which are well described by the resonant energy exchange mechanism of Campbell et al. \cite{campbell1983resonance}. Moreover, it has been shown that for $\epsilon\neq 0$, the resonance windows are suppressed because the normal modes became QNMs leading to energy loss. In this case more energy is lost during collisions and we are not able to give it back as tranlational energy afterwards. Therefore, the critical velocity is higher compared to the case with normal modes. These results are exactly matching with the ones in \cite{dorey2018resonant}. Finally, we characterized the fractal self-similarity of the resonance windows and the scaling behavior and showed that they are lost gradually as we set $\epsilon\neq 0$. This shows that normal modes are important for the system to exhibit self similarity in the kink and antikink collision.

\appendix

\section{Transmission and reflection coefficients}
\label{ap1}

The coefficients in eq. (\ref{coef}) are given by
\begin{align}
I_1=&\frac{J}{2}e^{k(L+\Delta L)}e^{im(L+\Delta L)}\beta^-,\nonumber\\
I_2=&\frac{J}{2}e^{-k(L+\Delta L)}e^{im(L+\Delta L)}\beta^+,\nonumber\\
H_1=& \frac{Je^{im(L+\Delta L)}}{2}\big\{e^{k\Delta L}\beta^-[\sin(pL)-\frac{k}{p}\cos(pL)]+e^{-k\Delta L}\beta^+[\sin(pL)+\frac{k}{p}\cos(pL)]\big\},\nonumber\\
H_2=& \frac{Je^{im(L+\Delta L)}}{2}\big\{e^{k\Delta L}\beta^-[\cos(pL)+\frac{k}{p}\sin(pL)]+e^{-k\Delta L}\beta^+[\cos(pL)-\frac{k}{p}\sin(pL)]\big\},\nonumber\\
G_1=& \frac{Je^{im(L+\Delta L)}}{2}e^{-kL}\big\{e^{k\Delta L}\beta^-[\cos(2pL)+\alpha_{kp}^-\sin(2pL)]-e^{-k\Delta L}\beta^+\alpha_{kp}^+\sin(2pL)\big\},\nonumber\\
G_2=& \frac{Je^{im(L+\Delta L)}}{2}e^{kL}\big\{e^{k\Delta L}\beta^-\alpha_{kp}^+\sin(2pL)+e^{-k\Delta L}\beta^+[\cos(2pL)-\alpha_{kp}^-\sin(2pL)]\big\},\nonumber\\
\begin{split}
F_1=& \frac{Je^{2im(L+\Delta L)}}{2}\big\{e^{2k\Delta L}[1+i\alpha_{km}^-][\cos(2pL)+\alpha_{kp}^-\sin(2pL)]-2i\alpha_{km}^+\alpha_{kp}^+\sin(2pL)\nonumber\\
  & +e^{-2k\Delta L}[1-i\alpha_{km}^-][\cos(2pL)-\alpha_{kp}^-\sin(2pL)]\big\},
\end{split}\nonumber\\	
\begin{split}
F_2=& \frac{J}{2}\big\{-ie^{2k\Delta L}\alpha_{km}^+[\cos(2pL)+\alpha_{kp}^-\sin(2pL)]+2i\alpha_{km}^-\alpha_{kp}^+\sin(2pL)\\
    &+ie^{-2k\Delta L}\alpha_{km}^+[\cos(2pL)-\alpha_{kp}^-\sin(2pL)]\big\},
\end{split}
\end{align}
using the definitions in eq. (\ref{defin}) and also
\begin{equation}
\beta^\pm \equiv 1\pm\frac{im}{k}\, .\nonumber
\end{equation}
The transmission and reflection coefficients are given by $T=|\frac{J}{F_1}|^2$ and $R=|\frac{F_2}{F_1}|^2$, respectively.

\section{Square well eigenfunctions}
\label{ap2}
The even eigenfunctions of the square well potential are given by
\begin{equation}
\eta_{2n}(x)= 
  \begin{cases} 
     C_{2n}e^{k_{2n}x}, & x<-L, \\
     C_{2n}e^{-k_{2n}L}\cos(p_{2n}x)/\cos(p_{2n}L), & -L<x<L, \\
     C_{2n}e^{-k_{2n}x}, & x>L,
  \end{cases}
\end{equation}
where $C_{2n}$ is the normalization constant. The parameters $k_{2n}$ and $p_{2n}$ are functions of $\omega_{2n}$, which is the $(n+1)$-th solution of the transcendental equation (\ref{normal1}) in increasing order. 

The odd eigenfunctions are given by
\begin{equation}
\label{eigenodd}
\eta_{2n+1}(x)= 
  \begin{cases} 
     -C_{2n+1}e^{k_{2n+1}x}, & x<-L, \\
     C_{2n+1}e^{-k_{2n+1}L}\sin(p_{2n+1}x)/\sin(p_{2n+1}L), & -L<x<L, \\
     C_{2n+1}e^{-k_{2n+1}x}, & x>L,
  \end{cases}
\end{equation}
with $C_{2n+1}$ as normalization constant. The parameters $k_{2n+1}$ and $p_{2n+1}$ are functions of $\omega_{2n+1}$, which is the $(n+1)$-th solution of the transcendental equation (\ref{normal2}) in ascending order as in the previous case.

\section{Numerical technique}
\label{ap3}
Substituting the partial derivatives by finite differences, one can solve the field equations numerically.
We divide the spacetime in a grid with spacings $\tau=0.002$ and $h=0.01$. We define the field at the gridpoint $(x_i,t_j)$ as $\phi_{i,j}$. Therefore, the partial derivatives take the form
\begin{equation}
\frac{\partial^2 \phi_{i,j}}{\partial t^2}=\frac{\phi_{i,j+1}-2\phi_{i,j}+\phi_{i,j-1}}{\tau^2},
\label{finite1}
\end{equation}
and
\begin{equation}
\frac{\partial^2 \phi_{i,j}}{\partial x^2}=\frac{\phi_{i+1,j}-2\phi_{i,j}+\phi_{i-1,j}}{h^2}.
\label{finite2}
\end{equation}
We set the boundaries at $x=\pm 100$, with periodic boundary conditions. We chose this method because it is the simplest method that works and because higher order methods do not deal well with discontinuities. We checked the energy conservation of the method and we found conservation up to a relative error of the order $10^{-4}$ during the full evolution of the system.

\section*{Acknowledgments}

We acknowledge financial support from the Brazilian agencies CAPES and CNPq. AM also thanks financial support from Universidade Federal de Pernambuco Edital Qualis A.

\end{document}